# Single-crystal and polycrystalline diamond erosion studies in Pilot-PSI

D. Kogut[1], D. Aussems[2], N. Ning[1], K. Bystrov[2], A. Gicquel[3], J. Achard[3], O. Brinza[3], Y. Addab[1], C. Martin[1], C. Pardanaud[1], S. Khrapak[1] and G. Cartry[1]

[1]*Aix-Marseille Université, CNRS, PIIM, UMR 7345, 13397 Marseille, France*
[2]*DIFFER - Dutch Institute for Fundamental Energy Research, De Zaale 20, 5612 AJ Eindhoven, Netherlands*
[3]*LSPM, Université Paris 13, Sorbonne Paris Cité, CNRS, 93430 Villetaneuse, France*

## Abstract

Diamond is a promising candidate for enhancing the negative-ion surface production in the ion sources for neutral injection in fusion reactors; hence evaluation of its reactivity towards hydrogen plasma is of high importance. High quality PECVD single crystal and polycrystalline diamond samples were exposed in Pilot-PSI with the $D^+$ flux of $(4–7)\cdot10^{24}$ $m^{-2}s^{-1}$ and the impact energy of 7–9 eV per deuteron at different surface temperatures; under such conditions physical sputtering is negligible, however chemical sputtering is important. Net chemical sputtering yield $Y = 9.7\cdot10^{-3}$ at/ion at 800°C was precisely measured *ex-situ* using a protective platinum mask (5x10x2 µm) deposited beforehand on a single crystal followed by the *post-mortem* analysis using Transmission Electron Microscopy (TEM). The structural properties of the exposed diamond surface were analyzed by Raman spectroscopy and X-ray Photoelectron Spectroscopy (XPS). Gross chemical sputtering yields were determined *in-situ* by means of optical emission spectroscopy of the molecular CH A-X band for several surface temperatures. We observed a bell shape dependence of the erosion yield versus temperature between 400°C and 1200°C, with a maximum yield of ~$1.5\cdot10^{-2}$ at/ion attained at 900°C. The yields obtained for diamond are relatively high $(0.5–1.5)\cdot10^{-2}$ at/ion, comparable with those of graphite. XPS analyses show amorphization of diamond surface within 1 nm depth, in good agreement with molecular dynamics (MD) simulation. MD was also applied to study the hydrogen impact energy threshold for erosion of [100] diamond surface at different temperatures.

Corresponding author: gilles.cartry@univ-amu.fr

# 1. Introduction

Deuterium negative ion generation is of a primary interest for fusion reactors as an essential component of the Neutral Beam Injection (NBI) system. The standard solution to produce high negative-ion current is to inject cesium (Cs) onto extraction grid, which has certain drawbacks for the long-term operation, such as Cs diffusion in the accelerator stage of NBI system and plasma contamination; hence, the development of cesium-free negative-ion sources is a major issue for future fusion reactors. It has been shown that diamond is a good negative-ion surface-production enhancer material when exposed to a low pressure hydrogen plasma [1,[2]. In negative-ion sources for fusion the plasma grid on which negative-ions are formed is biased between the floating and the plasma potential, therefore most of positive ions impinges on the grid with a quite low energy of few eV. Under such plasma conditions the flux of positive ions is important while their energy remains in the eV range. In view of the use of diamond for fusion applications such as negative-ion sources, a proper evaluation of diamond erosion under high flux of ions at low energy is proposed here.

Diamond has been barely studied as a plasma facing material (PFM) [3,[4,[5,[6]. It has outstanding thermal properties; the chemical erosion rate of diamond by thermal hydrogen atoms is 2-4 orders of magnitude lower than that of graphite depending on the surface temperature [7], although in case of 200−800 eV hydrogen ions the physical sputtering yields of diamond and graphite are comparable [8]. The previous studies in the linear plasma device Pilot-PSI have shown that diamond could be a suitable PFM for a fusion reactor [3]. Unfortunately, these measurements have lacked the control of the surface temperature during the exposure and only polycrystalline diamond samples have been investigated. In addition, the spectroscopic measurements of the erosion yield have not been absolutely calibrated. Therefore, a new revised experiment is described here. This study is focused on the damage and erosion of diamonds film during interaction with hydrogen plasma under experimental conditions that demonstrate high ion-flux (~$5.10^{24}$ $m^{-2}s^{-1}$) and low energy (~ 7-9 eV). The goal is to add value to the knowledge on diamond behavior under exposure to high-flux low-energy plasmas through the control of the surface temperature during measurements, through measurement of the absolute erosion yields and through the use of high quality PECVD diamond film (both polycrystalline and single crystals with different orientations). We also used molecular dynamics (MD) study to have an insight of erosion process under similar conditions.

In Section 2 the processes involved in chemical sputtering of carbon and diamond are reviewed; Section 3 presents experimental set-up and diagnostics. Measurements of the diamond erosion rate are given in Section 4 together with molecular dynamics modelling.

# 2. Chemical sputtering of carbon and diamond

Erosion of carbon has been widely studied in view of its potential application as a plasma-facing material in fusion reactors, in particular under well-controlled conditions using low-flux ion beams, as well as in high-flux experiments in tokamaks and linear plasma devices [9]. Thermally activated chemical erosion occurs through hydrogenation of carbon atoms on the surface, formation of intermediate $sp^x$ centers with dangling bonds and subsequent release of mainly CH$_3$ radicals [10,[11,[12]. This process is hindered at elevated temperatures by hydrogen recombination on the surface. As a result, a typical dependence of the chemical erosion yield of carbon $Y_{therm}$ on the surface temperature $T_{surf}$ is a bell-shaped curve with a maximum at $T_{max}$ [9]. The chemical erosion rate is enhanced if the surface is amorphous or if it is damaged by impinging ions, which leads to formation of dangling bonds for hydrogen attachment. The latter requires ion energy above a certain damage threshold. It has been also shown that $T_{max}$ shifts to higher temperatures and $Y_{therm}$ gradually decreases with the ion flux increase [9].

The total chemical sputtering yield of carbon, $Y_{tot}$, includes physical sputtering, chemical erosion $Y_{therm}$ enhanced by damage production and the near-surface process, $Y_{surf}$. The latter corresponds to sputtering of sp$^3$ CH$_n$ groups from the surface via breaking the weakly bound C-C bonds by ion impacts. The process itself is not thermally activated, however $Y_{surf}$ strongly depends on the surface density of sp$^3$ hydrocarbons which decreases at high $T_{surf}$ due to the dominance of hydrogen recombination and desorption.

A number of $Y_{tot}(T_{surf})$ curves have been calculated with a model of Roth [12] for several fluxes of deuterons impinging on graphite with energy of 8 eV, see Figure 1. Each curve has a bell-shaped contribution $Y_{therm}$ and a sigmoid-like contribution $Y_{surf}$. One could expect the same type of behavior for diamond if the topmost layer of its surface is amorphized under ionic bombardment.

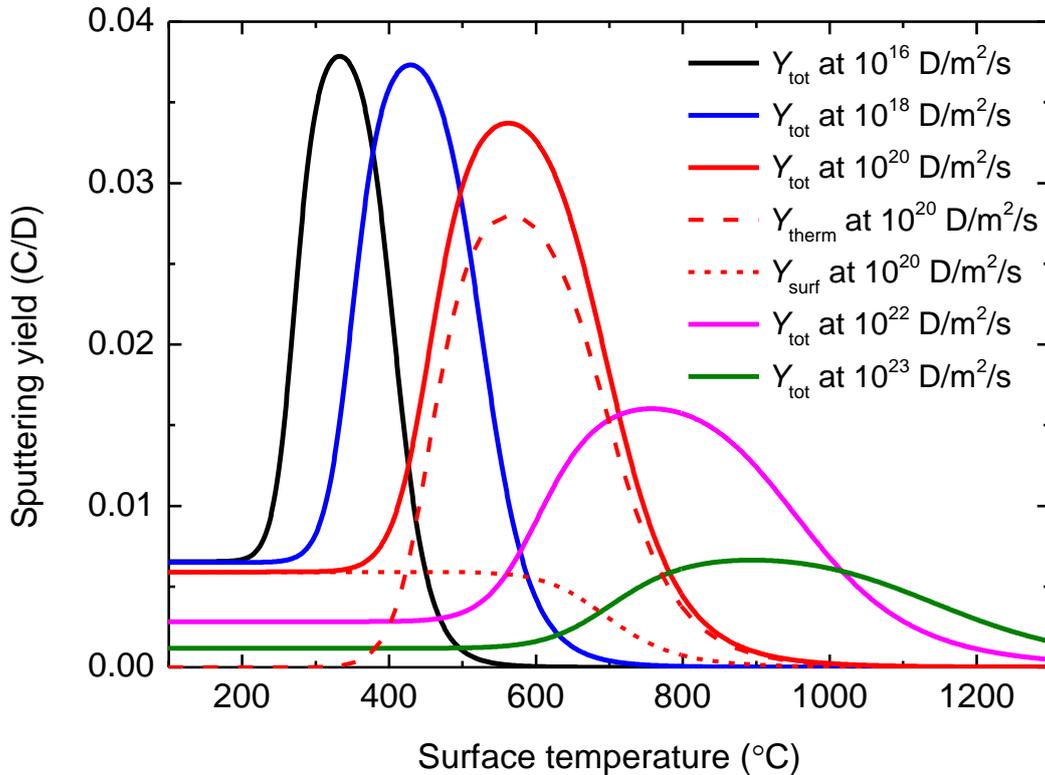

Figure 1. Total chemical sputtering yield $Y_{tot}$ of graphite under 8 eV D$^+$ ion bombardment, calculated with the model of Roth [12] as a function of the surface temperature for different fluxes of incident ions. $Y_{chem}$ is thermally activated chemical erosion yield and $Y_{surf}$ is the yield of near-surface sputtering of weakly bound sp$^3$ CH$_n$ groups.

Experimental evidence of chemical erosion of diamond under hydrogen or deuterium impact at low energy (few eV or less) is limited. Donnely has shown that a polycrystalline CVD diamond film is quite resistant under the flux of thermal H atoms of $8.9 \cdot 10^{21}$ m$^{-2}$s$^{-1}$ with erosion yield of the order of $10^{-7}$-$10^{-6}$ C/H with a slight increase towards $T_{surf}$ = 1100 K, while graphite under the same conditions revealed a bell-shaped curve with $Y_{max} = 3 \cdot 10^{-3}$ C/H at $T_{max}$ = 800 K [7]. Takeguchi studied graphite erosion in a high power Ar + $H_2$ ICP plasma with a low-energy neutral flux of $10^{23}$–$10^{24}$ m$^{-2}$s$^{-1}$ and an ion flux of $10^{19}$–$10^{20}$ m$^{-2}$s$^{-1}$ with incident energies of few eV; he found that deposition of boron-doped diamond film on top of graphite reduces its chemical sputtering yield at 850 K by 2 orders of magnitude down to $3 \cdot 10^{-5}$ C/H [13[20]. He also observed a significant modification of the surface morphology (formation of pyramidal pits) and H penetration 20 nm deep in the diamond layer. In previous experiments on Pilot-PSI the nano-crystalline and microcrystalline diamond CVD films were exposed to high flux hydrogen plasma: electron density $n_e = 2 \times 10^{19}$ m$^{-3}$, electron temperature $T_e$ = 0.3–1.4 eV (giving energy below 10 eV), ion flux of $2 \times 10^{23}$ m$^{-2}$s$^{-1}$. CH emission produced by erosion of diamond films was found to be reduced by a factor 2 compared to graphite if $T_e$ was less than 1 eV [3]; unfortunately, $T_{surf}$ could not be measured at that time. Partial amorphization of the diamond structure within the penetration depth of ions (appearance of sp$^2$ carbon in XPS) occurred at $T_e$ = 1.5 eV [3]. When dealing with higher ion energy, above ~10 eV, more studies devoted to erosion of diamond and defect creation on diamond can be found. Yudo et al [14] observed in CVD reactor that the size of diamond particles were reduced to one third when the bias was set to -100 V and the growth of diamond crystal was completely suppressed when the bias was set to -200 V, suggesting that the effect of sputtering and erosion by hydrogen is very large at high ion enernngy. V. Yamada used high energy $H_2^+$ ion beam of $2 \cdot 10^{15}$ m$^{-2}$s$^{-1}$ with 200–800 eV/H and he observed similar chemical sputtering yields both for graphite, sintered diamond and diamond film, of the order of $10^{-2}$-$10^{-1}$ C/H, depending on the ion energy; he also evidenced a bell-shaped curve $Y(T_{surf})$ with $T_{max}$ around 500°C [8]. Yamazaki exposed high-pressure and high-temperature (HPHT) synthetic [001] diamond substrates to 50 W RF plasma discharge at 10 Pa of $H_2$, with ion energies up to 500 eV at room temperature; subsequent XPS and FTIR analysis showed that such exposure leads to a structural change of a diamond structure towards an *a*-C:H-like one within 10 nm depth [21]. Microcrystalline CVD diamond films were exposed to a hydrogen ion flux of $10^{20}$ m$^{-2}$s$^{-1}$ in the linear magnetized plasma device MAGPIE with target biases between 0 V to −500 V leading to ion energy in the range 18 eV to ~500 eV; the ion-induced damage to the surface of the diamond films occurred only within the ion penetration depth (10-15 nm), in agreement with the SRIM modelling [5]. After exposure to $10^{23}$ deuterons per m$^2$ in MAGPIE a lot of 10 nm–100 nm sized hemispherical and conical features appeared on the CVD surface, which was explained by re-deposition of the eroded carbon [5[6]. Surface analysis of CVD diamond films exposed in tokamak DIII-D showed that the bulk micro-crystalline nature of the sample is unaltered while a continuous 10-15 nm thick interfacial layer was formed at the surface; a structural composition of 24% diamond and 76% amorphous carbon in the layer was measured [4]. Finally, many other studies have also been devoted to damage creation in diamond upon exposure to heavier ions (carbon, oxygen, argon, xenon, …) in a wide range of impact energies [15[16[17[18[19].

To summarize the experiments listed above, it is expected that the chemical sputtering of diamond occurs through amorphization of the sub-surface layer within the ion penetration depth. It is also suggested that there is a certain ion energy threshold to create the damaged layer; that is why diamond erosion by thermal H atoms was found to be negligible in contrast to the case of exposure to energetic hydrogen or deuterium ions. Indeed, the displacement energy of a carbon atom in diamond lattice is 52 eV [22]; in order to transfer such energy to a C atom through an elastic collision, an H atom must have at least 183 eV [5]. In case of graphite an ion energy of 15–30 eV is required to displace a C atom permanently [23], which corresponds to an H atom with a minimum energy of 53–106 eV [5]. On the other hand, even low-energy H ions can cause erosion of amorphous carbon and graphite via breaking C–C bonds in the near-surface process; the latter requires approximately 2–8 eV, depending on the bonding configuration [24]. Previous exposures of fine-grain graphite to high flux hydrogen plasmas in Pilot-PSI showed the threshold energy of chemical sputtering of 1.1 eV obtained by fitting the experimental data [25]. The near-surface sputtering occurs on a timescale of ps, hence it can be perfectly simulated with molecular dynamics (MD) approach. On the other hand, MD cannot describe long-term thermally activated processes, such as diffusion and desorption of hydrogen. MD simulation results will be presented in the paper together with experimental results.

## 3. Experiment

Diamond samples have been exposed in Pilot-PSI in order to measure the erosion rate and to study the modification of the surface properties induced by particle bombardment. Two different types of diamond layers were prepared by means of plasma-assisted chemical vapour deposition (PACVD). First one is a single crystal, either 3x3 mm² with [100] orientation or 2x2 mm² with [111] orientation, deposited on a low quality HPHT diamond substrate which in turn was brazed on a molybdenum substrate. The thickness of the crystal ranged from 20 to 64 μm, some of them were boron-doped with $10^{19}$–$10^{20}$ B/cm$^3$ (see Table 1). The second type of sample is a polycrystalline layer with a grain size of 20−100 μm and a thickness of around 100 μm, brazed on a molybdenum substrate. One of the polycrystalline samples was boron-doped with $10^{21}$ B/cm$^3$; the shape of its crystallites is slightly different compared to non-doped samples. The polycrystalline samples were circular with a diameter of 10 mm, see Table 1.

|  | Polycrystalline | | Single crystal | | | | | |
| --- | --- | --- | --- | --- | --- | --- | --- |
|  | | | [100] | | | [111] | |
|  | B-doped | Non-doped | B-doped | | | B-doped | Non-doped |
| **Reference** | MCD1 | MCD2 | MCD3 | SCD1 | SCD2 | SCD3 | SCD4 | SCD5 |
| **Thickness** | 100 μm | 100 μm | 100 μm | 20 μm | 20 μm | 20 μm | 64 μm | 22 μm |
| **Dimensions** | Ø10 mm | Ø10 mm | Ø10 mm | 3×3 mm$^2$ | 3×3 mm$^2$ | 3×3 mm$^2$ | 2×2 mm$^2$ | 2×2 mm$^2$ |
| $T_e$ (eV) | 1.5 | 1.4 | 1.4 | 1.4 | 1.7 | 1.1 | 1.5 | 1.7 |
| $n_e$ (m$^{-3}$) | 7.5·10$^{20}$ | 9.0·10$^{20}$ | 9.0·10$^{20}$ | 6.0·10$^{20}$ | 7.0·10$^{20}$ | 6.0·10$^{20}$ | 6.0·10$^{20}$ | 8.0·10$^{20}$ |
| D$^+$ flux (D/m$^2$/s) | 5.2·10$^{24}$ | 6.0·10$^{24}$ | 6.0·10$^{24}$ | 4.0·10$^{24}$ | 5.2·10$^{24}$ | 3.6·10$^{24}$ | 4.2·10$^{24}$ | 5.9·10$^{24}$ |
| D$^+$ energy (eV) | 7.8 | 7.3 | 7.3 | 7.3 | 8.8 | 5.7 | 7.8 | 8.8 |
| Exposure time (s) | 10 | 10 | 40 | 10 | 10 | 10 | 40 | 10 |
| D$^+$ fluence (D/m$^2$) | 5.2·10$^{25}$ | 6.0·10$^{25}$ | 2.4·10$^{26}$ | 4.0·10$^{25}$ | 5.2·10$^{25}$ | 3.6·10$^{25}$ | 1.7·10$^{26}$ | 5.9·10$^{25}$ |
| Surface temp. (°C) | 890 | >900 | 1200 | 800 | 730 | 380 | 500 | 550 |

Table 1. Properties of the exposed diamond samples and the parameters of D$_2$ plasma exposure in Pilot-PSI.

The linear plasma device Pilot-PSI (Figure 2) has been chosen to expose the diamond samples as it provides a huge particle flux at low energy of impact and low electron temperature of D$_2$ plasma [26]. Plasma generated by a cascaded arc source [27] expands towards the target along a magnetic field of 0.8 T. The source current and the deuterium flow were adjusted to obtain the reproducible plasma parameters: the electron temperature $T_e$ of 1.4–1.7 eV, the electron density $n_e$ of (0.6–1.0)·10$^{21}$ m$^{-3}$ as measured by Thomson scattering in front of the target. The D$^+$ flux $\Gamma_i$ was calculated from the generalized Bohm criterion assuming $T_e = T_i$ [28]:

$$\Gamma_i = 0.5 n_e [(1 + \gamma) k_B T_e / M_i]. \tag{1}$$

$M_i$ is the ion mass, $k_B$ is the Boltzmann constant and $\gamma$ is assumed to be 5/3 (adiabatic flow with isotropic pressure). The target was at floating potential, hence the ion impact energy was 7–9 eV per deuteron. Neutrals have 10 times lower energy than the ions since they are not accelerated in the sheath. Their flux is supposed to be much lower than the ion flux since the ionization degree is very high in Pilot PSI (85% measured in [29]). Therefore neutrals are expected to contribute much less to the erosion. The neutrals were also neglected in previously published papers about Pilot-PSI, e.g. in [25]. The experimental conditions are close to those predicted for the ITER divertor plasma close to the strike point. Under such conditions physical sputtering is negligible, however chemical sputtering is important. The chemical sputtering yield strongly depends on the surface temperature, hence different surface temperatures and fluences were tested with the sample of each type. The target in Pilot-PSI is water-cooled; in order to vary the surface temperature a different number of graphite foil layers were introduced between the sample and the target for a fixed plasma condition. The temperature was measured during the exposure by means of a multi-wavelength pyrometer (FAR Associates, FMPI SpectroPyrometer) and a fast infrared camera (FLIR, SC7500-MB).

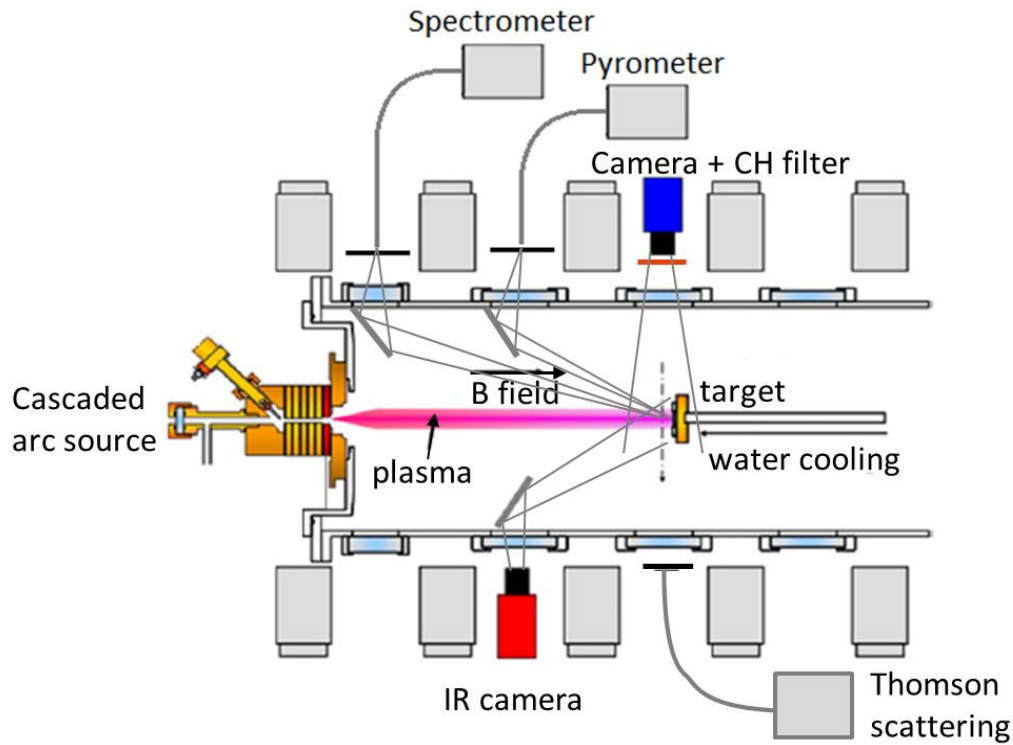

Figure 2. Scheme of the experiment in Pilot-PSI.

Gross chemical sputtering yield was determined *in-situ* by means of optical emission spectroscopy of the molecular CH A-X band (431.42 nm) [30]. The line was measured with Avantes AvaSpec-2048 spectrometer as well as with fast visible-range Phantom camera with a bandpass filter from 430.0 to 431.5 nm. Photon efficiency was calibrated by means of the methane ($CH_4$) injection close to the target without sample during the plasma pulse with the same conditions.

Single crystal [100] sample SCD1 was coated with a tiny platinum mask ($5\times10\times2$ $\mu m^3$) before exposure using a focused ion beam (FIB) setup HELIOS 600 Nanolab. After exposure FIB was applied again to cut a thin foil cross-section including exposed and unexposed diamond surfaces; the differential height between both surfaces was accurately measured by means of transmission electron microscopy (TEM) on FEI Titan 80300. Surface properties of the samples were studied with scanning electron microscopy (SEM) using a Philips XL30 SFEG, atomic force microscopy (AFM) using a NT-MDT apparatus, Raman microspectrometer Horiba-Jobin-Yvon HR LabRAM using $\lambda = 633$ nm and X-ray photoelectron spectroscopy (XPS) with Al-$K_\alpha$ source (1486.7 eV).

## 4. Results and discussion

### *4.1. Measurements of the chemical sputtering rate*

SEM images of the platinum mask in the center of the SCD1 sample before and after exposure to $4.0\cdot10^{25}$ $D^+/m^2$ at $T_{surf} = 800$°C during 10 s are shown in Figure 3. The mask protected the diamond from erosion, its rectangular shape became a bit smeared probably due to a localized heating and melting of platinum; physical sputtering of Pt by deuterons with 7 eV is not possible. Small droplets of Pt were deposited on the distance of ≥10 μm around the mask during the exposure, see Figure 3b. The area around the mask was coated with another layer of platinum to protect the diamond surface and a thin foil cross-section was cut with FIB, which was further analyzed with TEM (Figure 4).

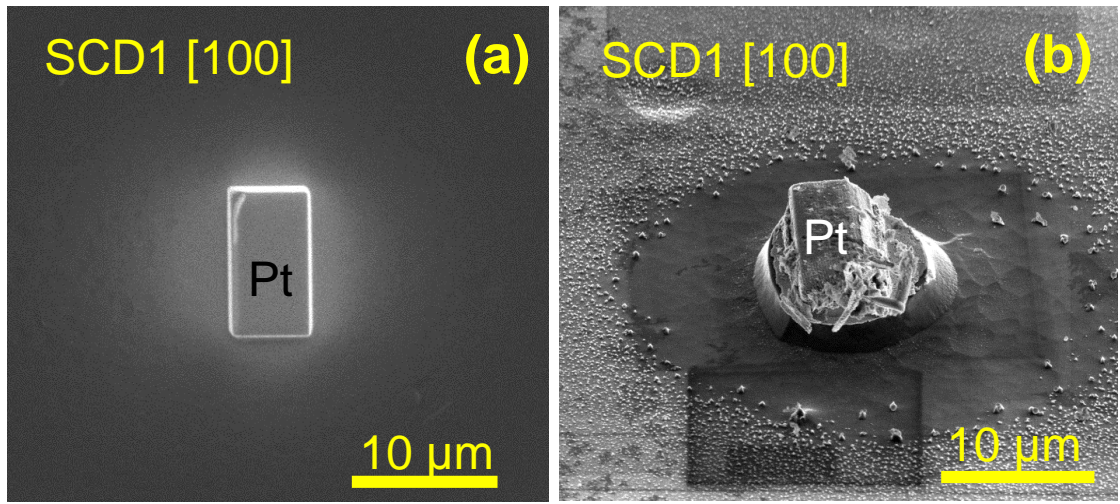

Figure 3. SEM image of the single crystal [100] diamond surface of the SCD1 sample with a Pt mask on top before (a) and after exposure (b) in Pilot-PSI. The image (b) was taken at the tilt angle of 52°.

On the cross-section a light-grey layer on top is platinum, a dark-grey layer on bottom is diamond. On the right the diamond [100] surface was protected by the mask during the exposure in Pilot-PSI: a clear erosion step can be observed with respect to the unprotected area on the left. The height of the step is $(2.2 \pm 0.2)$ μm, which yields in the sputtering rate of $(220 \pm 20)$ nm/s. Given the $D^+$ flux to the surface of $4.0 \cdot 10^{24}$ $m^{-2}s^{-1}$ and the diamond density of 3.5 g/cm$^3$, the chemical sputtering yield of [100] surface bombarded by 7 eV deuterons at 800°C is $Y = (9.7 \pm 0.9) \cdot 10^{-3}$ at/ion. The interface between the coated platinum layer and the diamond substrate was investigated in detail by means of high-resolution TEM, see Figure 5.

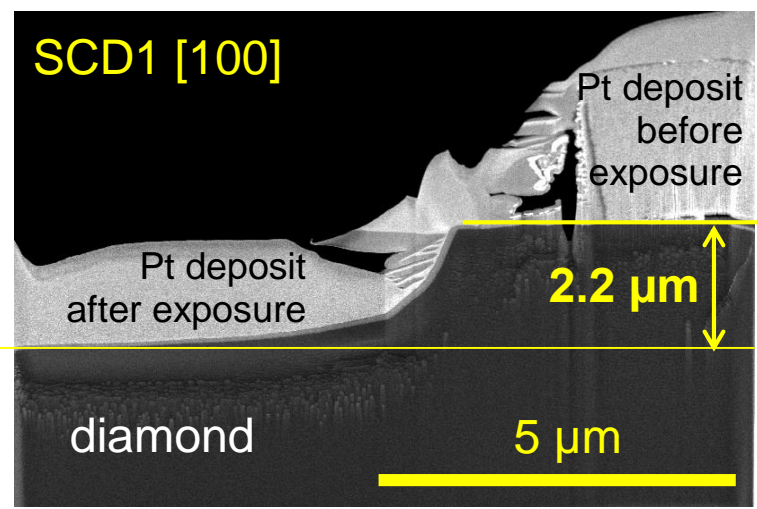

Figure 4. TEM image of the thin foil cross-section of the SCD1 sample, showing the platinum mask and the underlying diamond surface after exposure in Pilot-PSI. The erosion step is highlighted with yellow lines.

The only difference between the diamond layer protected with Pt mask during plasma exposure (Figure 5a) and non-protected area, covered with Pt after exposure (Figure 5b) is a light-grey interface layer of 1-2 nm thickness in the latter case. This could be due to amorphization of the diamond surface during plasma exposure. Raman spectroscopy showed a well-pronounced diamond peak before and after exposure with no signature of amorphous C in the spectra, which may be explained by its limited sensitivity: Raman scattering originates from the probing depth of a few hundreds of nm, while the amorphous layer thickness is possibly of the order of 1 nm.

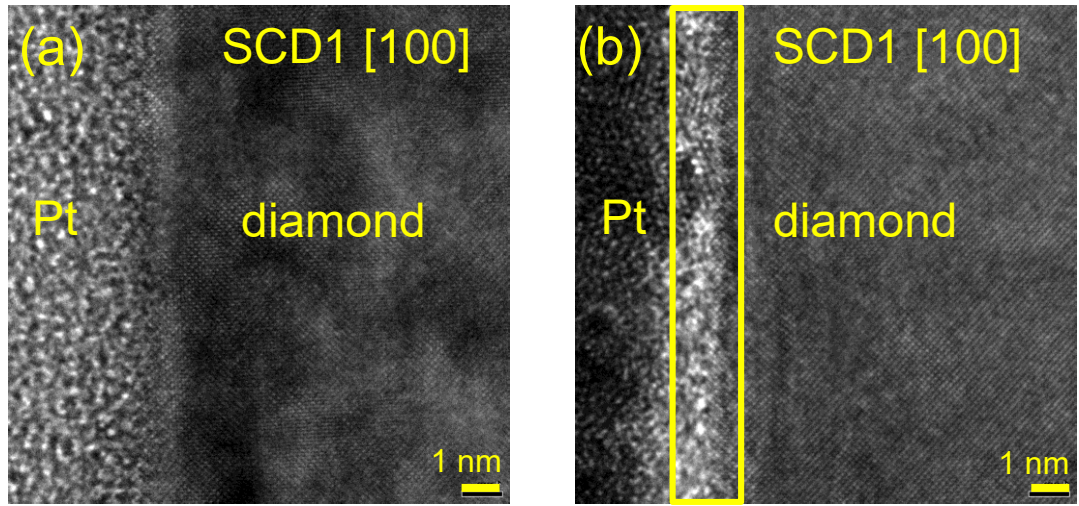

Figure 5. High-resolution TEM image of the interface between the platinum layer and the underlying diamond surface of the SCD1 sample: (a) masked area, (b) area exposed in Pilot-PSI. The yellow rectangle shows a light-grey interface layer which is probably due to the local amorphization of the diamond surface.

On the other hand, XPS spectra reveal a certain change of C1s line shape between unexposed [100] sample (Figure 6a) and the exposed SCD1 (Figure 6b). The spectrum was fitted with four Gaussian/Lorentzian peaks with 70% Gaussian contribution each; the major peak corresponds to $sp^3$ binding, it is located between 284.2 and 285.5 eV depending on the sample. Such variation can be firstly due to the charging of the diamond surface during the XPS analysis, though it is less probable as most of the samples are boron-doped. Another reason is that due to the 5.5 eV band gap in diamond the presence of surface states or dopants determines the C1s binding energy, which can range from 283.7 to 289.2 eV [32]. The difference of doping levels across the samples could explain the shift of the Fermi level, hence the shift of the binding energy peaks. Another recent review about the origins of $sp^3$ peaks in C1s XPS of carbon materials shows that the binding energy reported for $sp^3$ in the literature ranges from 283.3 to 287 eV [33]. However the shift between $sp^2$ and $sp^3$ peaks is reproducible by different experimenters: 0.7−1.1 eV [33], so we adopted the commonly used approach [34]: the absolute binding energies were allowed to change from sample to sample, however the relative shift between $sp^2$ and $sp^3$ peaks positions was fixed to 0.7−0.8 eV. FWHM for both peaks was found to be in the range of 0.5−1.1 eV.

While the shape of the diamond $sp^3$ peak remains unmodified for both samples in Figure 6, the peak of $sp^2$ is more pronounced for the exposed sample; $sp^2/sp^3$ ratio raised from 0.018 to 0.34 after exposure. Inelastic mean-free-path of 1.2 keV photoelectrons in amorphous carbon is 2 nm [31], which means that 65% of the measured signal originates from the topmost 2 nm layer and 95% within the depth of 6 nm: therefore, XPS provides more precise measurement of the surface composition compared to Raman in our case.

| Sample | C1s: $sp^2/sp^3$ | O1s |
|---|---|---|
| **SCD4: [111] at 500°C** | 0.12 ± 0.01 | 9.8 at.% |
| **SCD1: [100] at 800°C** | 0.34 ± 0.02 | 8.7 at.% |
| **MCD3: polycrystalline at 1200°C** | 0.043 ± 0.003 | 6.6 at.% |
| **[100] unexposed** | 0.018 ± 0.002 | 3.0 at.% |

Table 2. The ratio of $sp^2$ to $sp^3$ hybridization of C atoms and the quantity of oxygen measured by XPS.

XPS also shows a non-negligible amount of oxygen on the surface (3-10 at.%) which is higher for exposed samples. Besides, there are two small broad peaks in C1s line corresponding to C–O and C=O binding energies (Figure 6). Unfortunately the analysis was not performed immediately after the exposure in Pilot-PSI, which makes it impossible to deduce if the presence of oxygen in the vacuum chamber (base pressure $10^{-2}$ mbar) enhanced the chemical erosion or it was adsorbed later *ex-situ*.

It has been demonstrated previously in Pilot-PSI that in case of graphite targets there are erosion dominated zones and re-deposition dominated zones [35]. Under the present experimental conditions we believe that the sample is within the erosion dominated zone, hence the amorphous layer on the diamond surface is not a deposited C:D layer. Indeed, if the temperature at the center of the target corresponds to the maximum erosion rate, then erosion is dominating, while for higher temperatures it is re-deposition dominated. The 3 x 3 mm$^2$ diamond sample ABJ44 was exposed at $T_{surf} = 800$°C and demonstrated the sputtering yield close to the maximum value (see Figure 7 below), so we expect the sample to be in the erosion dominated zone (at least the analyzed area with the Pt mask).

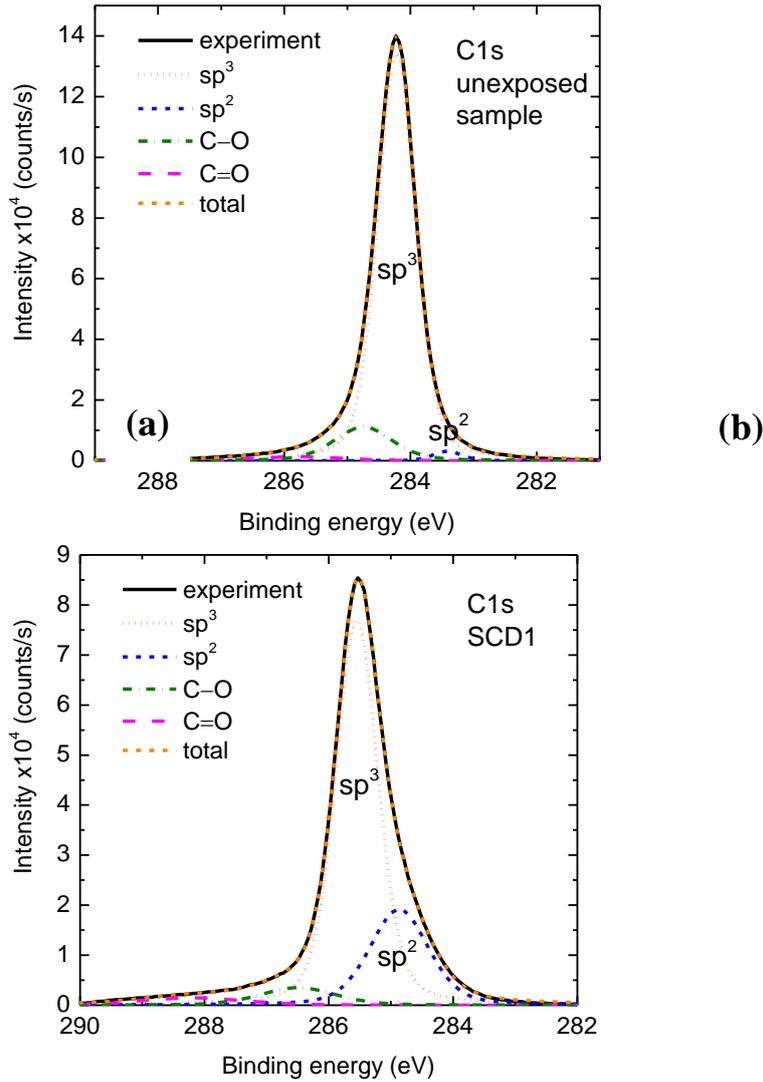

Figure 6. XPS spectra of C1s line of single crystal [100] samples: unexposed (a) and exposed SCD1 (b).

To conclude, the experimental evidence suggests that amorphization of the diamond surface could occur within 1-2 nm depth. Calculations with the SRIM code [36] predict that the ion penetration depth lies within 1 nm in our conditions. However, SRIM model assumes amorphous carbon target and it does not work correctly in the range of energies lower than 10 eV, so MD modelling was performed to account for the diamond [100] structure, as it will be shown in Section 4.2.

The measurements of the CH emission during plasma pulses for all samples allowed to obtain gross chemical sputtering yield as a function of the surface temperature and the crystalline structure, see Figure 7. The yield follows a bell-shaped curve mentioned in Section 2 with $T_{max}$ around 900°C. It should be noted that all measurements with $T_{surf} < 900$°C were performed with single crystal diamonds, while polycrystalline samples were exposed at $T_{surf} \geq 900$°C. The choice was determined by the type of Mo substrate and the type of diamond crystal brazing. Furthermore, $T_{surf}$ depended on exact heat flux for a given plasma pulse and thermal contact with the target, which made it hard to control.

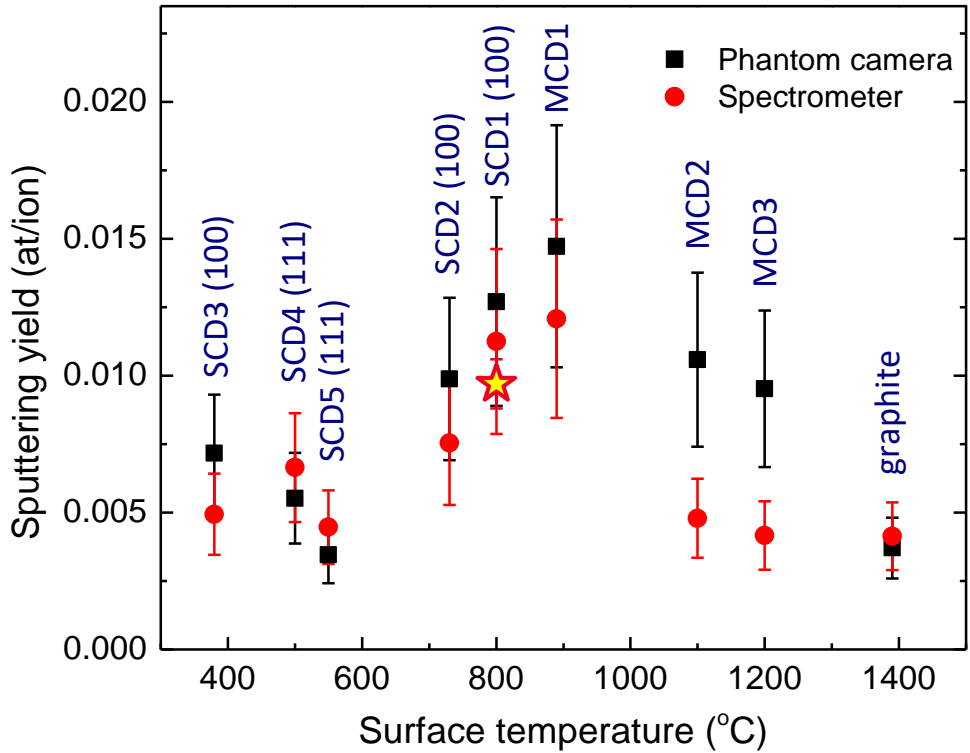

Figure 7. Gross chemical sputtering yield of diamond deduced from the CH emission measured *in-situ* with fast camera (black squares) and spectrometer (red circles) as a function of $T_{surf}$ and crystalline orientation. The star shows net sputtering yield measured *ex-situ* by surface analysis technique for one of the samples.

There is an uncertainty of the measurement related to the choice of the effective area on the diamond surface which is eroded and contributes to the CH emission; in case of fast camera there is also uncertainty about the integration area. The error bars in Figure 7 mostly stem from the ~30% uncertainty of estimation of the effective flux area: the sample is smaller than the ion beam with Gaussian distribution, so the integrated ion flux over the target area is not accurately known. Nevertheless, the erosion yields measured by fast camera and spectrometer are in a reasonable agreement for most of the samples except for polycrystalline MCD2 and MCD3. The gross erosion yield for SCD1 is slightly higher than the net sputtering yield measured by TEM for this sample, shown by a star in Figure 7 (though it lies within the error bars): it implies local re-deposition of up to 25% of eroded carbon. Two crystalline orientations [100] and [111] did not show much difference in the sputtering yields in the range of $T_{surf}$ = 400–550°C. It should be noted that [111] surface at 500°C evidenced less amorphization ($sp^2/sp^3$) than [100] at 800°C, as given in Table 2; in fact, this is probably not related to the crystalline orientation but rather to higher $T_{surf}$ in case of [100] surface.

### *4.2. Modelling of the diamond [100] surface sputtering*

Sputtering yields obtained for diamond are relatively high: $(0.5–1.5) \cdot 10^{-2}$ at/ion and comparable with the value measured for graphite at 1400°C: $0.4 \cdot 10^{-2}$ at/ion. Temmerman also observed similar levels of CH emission in case of diamond and graphite exposed to $D_2$ plasma at $T_e$ = 1.3 eV in Pilot-PSI [3]. This suggests that the sub-surface layer facing plasma flux is modified by impinging $D^+$ ions in a similar way for both diamond and graphite. In order to gain an insight into the diamond surface modification at these experimental conditions, we have carried out molecular dynamic (MD) simulations [37] based on a Tersoff–Brenner type reactive empirical bond order (REBO) potential for hydrocarbons [38]. The REBO potential has been originally developed for simulating the PECVD deposition of diamond films [39]. This potential has been also used to model particle surface reaction [40], fullerene formation and properties [41], diamond melting [42], carbon nanotube properties [43], and polycrystalline diamond structure [44]. This formalism takes into account covalent bond breaking and formation with associated changes in atomic hybridization, providing a powerful method for modeling complex chemistry in large many-body systems.

The simulated D-terminated diamond [100]-2×1 cell is 2.8 nm × 2.8 nm × 2.5 nm; its side view before exposure is shown in Figure 8. The modelling was performed for three surface temperatures: 427°C, 627°C and 827°C. Before simulating the ion impacts the modelling cell was pre-heated up to the given $T_{surf}$ in a constant-temperature, constant-pressure (NPT) ensemble, which was followed by a relaxation in a microcanonical ensemble for 20 ps. In order to better reproduce the erosion process at high surface temperature, the simulation cell was separated into four different zones: a rigid zone of fixed atoms at the bottom, a temperature controlled zone to efficiently cool down the surface during impacts, an impact zone and a buffer zone in a microcanonical ensemble during the ion impacts (Figure 8). MD simulation includes a series of $D^+$ impacts with impact energy of 8 eV; the cell is allowed to cool down and relax for 0.5 ps between each impact. The total number of impacts is chosen to be high enough (6500 impacts corresponding to the fluence of about $6·10^{21}$ $D/m^2$) to achieve a quasi-equilibrium state in terms of erosion rate and amorphization state ($sp^2/sp^3$ ratio).

Figure 9a-c demonstrate the atomic configuration of the simulated cell after exposure to a fluence of $6·10^{21}$ $D/m^2$ at different surface temperatures. Amorphization and formation of C:D chains was observed in the modified layer of the cell at $T_{surf}$ of 427°C and 627°C, while at 827°C the sub-surface region becomes graphitized. Indeed, diamond undergoes phase transition to thermodynamically stable graphite through a thermally induced process [45]; bond breaking on diamond surface upon annealing can also initiate graphitization process [46]. Dunn et al. [47] also found that graphitization of diamond can be induced by ion impacts at surface temperatures above 727°C.

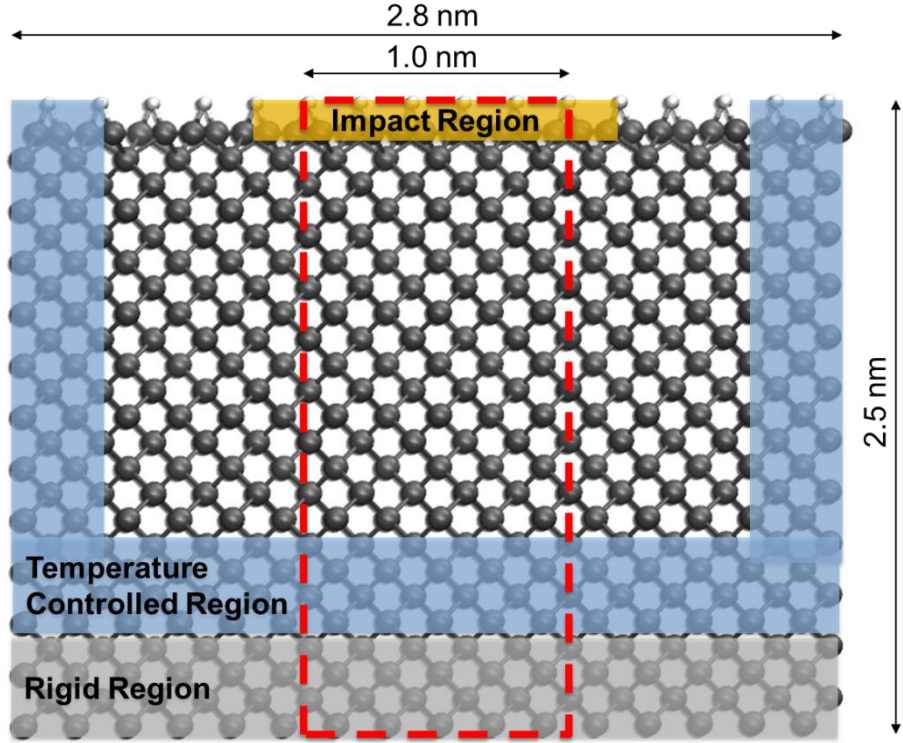

Figure 8. Atomic configuration of the simulated D-terminated C[100]-2×1 diamond cell (2.8 nm × 2.8 nm × 2.5 nm). The atoms inside gray region at the bottom are fixed during the simulation; the thermostat region is placed above, marked in blue. Incident $D^+$ ions bombard the surface within the impact region boundaries, shown in yellow. The rest of atoms between the impact zone and the thermostat zone are in the buffer region, where there is no temperature control. Red dash line defines the sampling region (1 nm × 1 nm × 2.5 nm) where atoms were sampled for the surface modification analysis.

The distribution of deuterium atoms as a function of the cell thickness given in Figure 9d confirms the composition change of the sub-surface region within 0.7–1.0 nm depth for three different surface temperatures. D/C ratio on top of the simulated cell is around 3 when $T_{surf} <$ 627 °C, which suggests formation of abundant $CD_3$ groups on the surface. At $T_{surf}$ = 827 °C D/C ratio in the top surface layer is about 2, which means that $CD_2$ groups are dominant.

The degree of amorphization can be evaluated through the $sp^2/sp^3$ ratio, which is shown as a function of $T_{surf}$ in Figure 10b. The $sp^2/sp^3$ ratio rises from 0.036 to 0.35 as $T_{surf}$ increases from 427°C to 827°C. These simulation results are in a good agreement with those given by the XPS measurements. Carbon sputtering yields obtained by MD simulations agree with the gross erosion yields measured in the experiment (see Figure 10a). Both sputtering yield and $sp^2/sp^3$ ratio were evaluated within the sampling region shown by red dashed line in Figure 8. The error bars in MD simulations were estimated based on series of three independent calculations in each case.

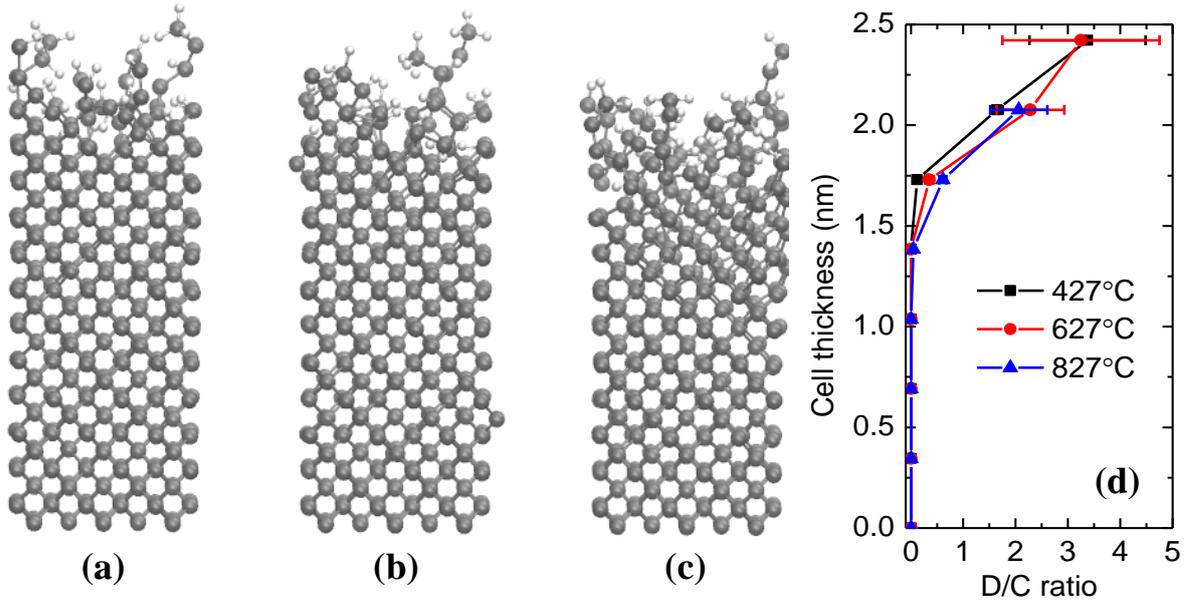

Figure 9. Atomic configuration of the simulated D-terminated C[100]-2×1 diamond cell after D impacts with energy of 8 eV and fluence of $6 \cdot 10^{21}$ D/m$^2$ at (a) 427°C, (b) 627°C, and (c) 827°C, respectively. Carbon atoms are shown as dark gray spheres and deuterium ones as small light gray spheres; (d) D/C ratio vs. the cell thickness after D impacts with different surface temperatures. The cell thickness of 2.5 nm corresponds to the topmost surface layer of the initial cell used in the calculations.

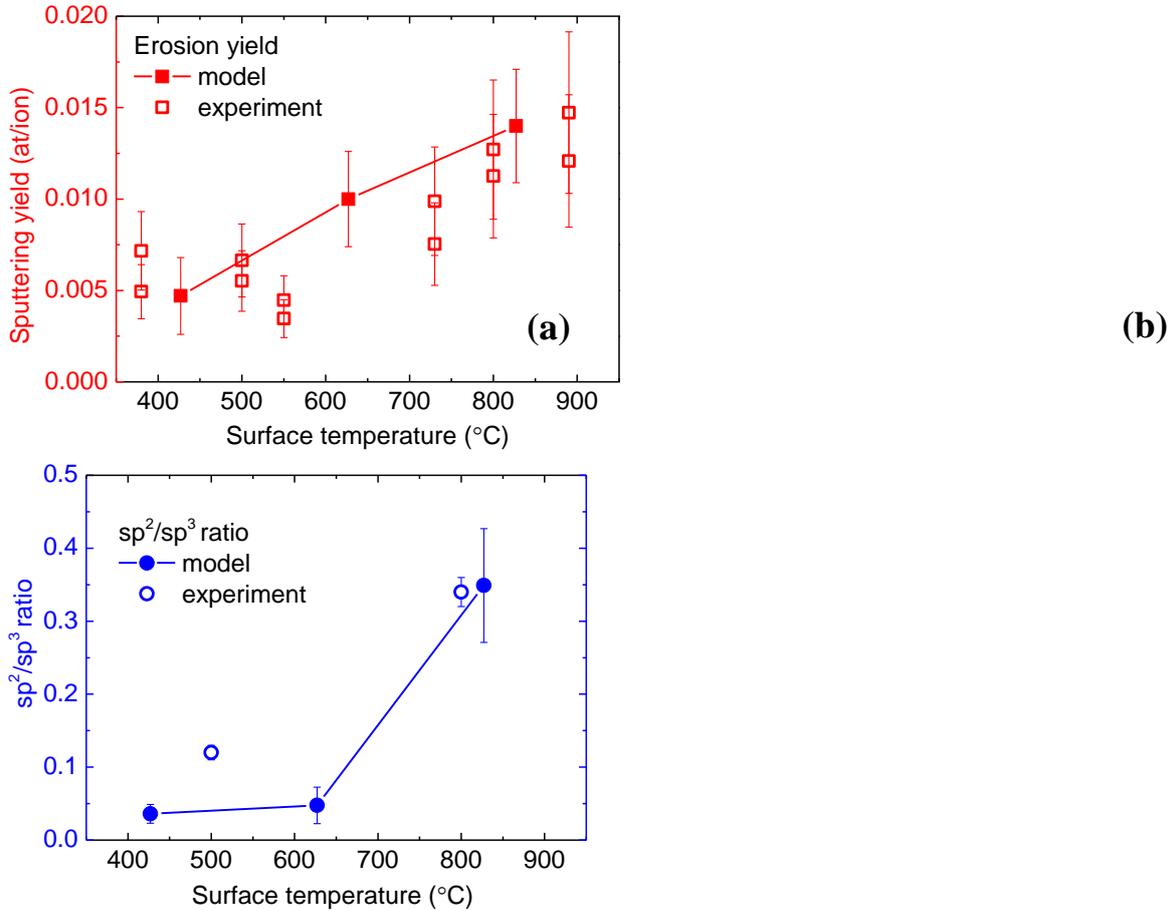

Figure 10. Carbon sputtering yield (a) and sp$^2$/sp$^3$ ratio (b) as a function of $T_{surf}$ given by the MD simulation and its comparison with experimental data.

Chemical sputtering of diamond by D impacts follows a layer-by-layer removal mechanism. The process begins with hydrogenation of the sub-surface region. As D atoms are incorporated into C-C bonds, the sub-surface region becomes amorphized and expands because of the creation of $CD_x$ chains. Weakly bound hydrocarbon molecules in the impact zone can be removed from the surface due to the C-C bond breaking via kinetic or thermally activated mechanisms; the latter mechanism occurs on a long-term timescale, therefore it is not included in the MD simulation. The binding energy between C atoms, hence chemical sputtering yield, is sensitive to the surface structure, namely to the number of neighbouring C, D atoms and bonds [48]. With the increase of $T_{surf}$ up to 827°C the fraction of $sp^2$ bonding in the sub-surface region is increased and $CD_2$ groups become dominant; consequently chemical sputtering yield of carbon may also increase. At $T_{surf} > 1000°C$ a lower surface concentration of D can be expected due to prevailing $D_2$ recombination and desorption, therefore chemical erosion yield decreases, which is observed in the experiment. It is not observed in the MD simulation as $D_2$ recombination can occur through Eley–Rideal or Langmuir–Hinshelwood mechanisms and the latter is not included in the model.

Given the synergetic effect of ion impacts and surface temperature, it is very important to understand the role of thermal effects when ion energy is low. Nakamura et al [49] studied the hydrogen injection to diamond surface with different impact energies (0.3 eV, 0.5 eV, 0.7 eV, 1 eV, 5 eV) using an unconventional ion impact method where the surface temperature was 0 K initially, and no cooling regime occurred after each impact (1750 impacts in total). It was found that carbon sputtering only occurred in the 5 eV impact energy case with a surface temperature saturated around 1227°C after 800 impacts. It is therefore interesting to examine the impact energy region below 5 eV with proper consideration of thermal effects. Using a conventional accumulative ion impacts method, MD simulation was employed to define energy threshold under conditions similar to our experiments. Figure 11 demonstrates the threshold behaviour of the chemical sputtering yield as a function of the D impact energy at room temperature calculated on a 1.4 nm × 1.4 nm × 2.5 nm C[100]-2×1 cell. The diamond [100] surface is hydrogenated and slightly modified with the sputtering yield lower than $2 \cdot 10^{-3}$ at/ion when $D^+$ energy is below 5 eV. This might be a promising result for the application of diamond layers in negative-ion sources for NBI where huge fluxes of low-energy deuterons on the convertor grid are expected. As $T_{surf}$ increases up to 827°C, the impact energy threshold decreases to 2 eV with a carbon sputtering yield of $1.7 \cdot 10^{-3}$ at/ion. In the future work, an optimal combination of impact energy, surface temperature and ion flux should be found in order to achieve maximum negative ion surface production with minimum surface damage.

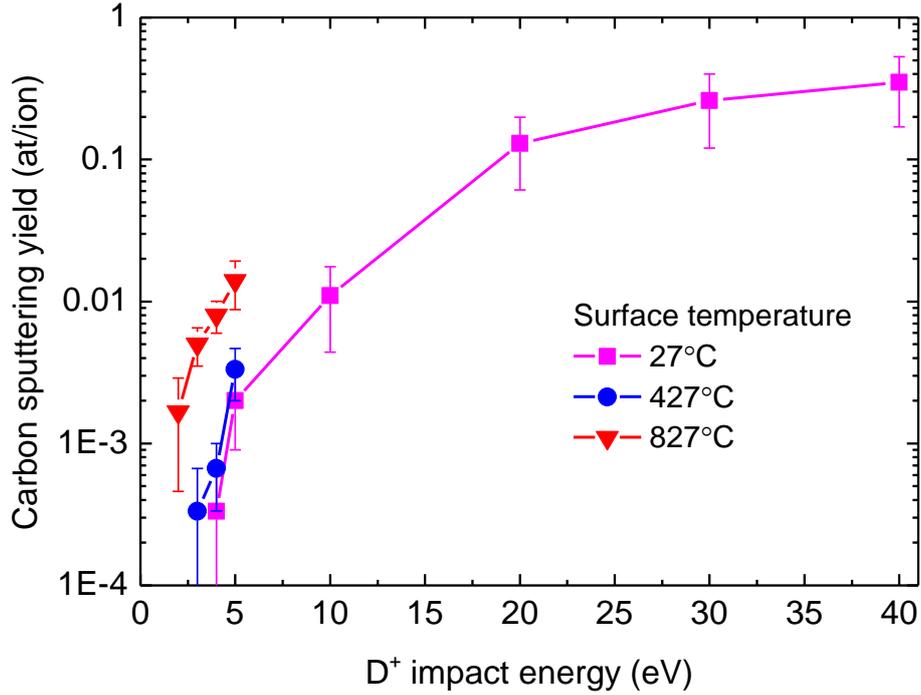

Figure 11. Impact energy dependence of the carbon sputtering yield at different $T_{surf}$ given by MD simulation.

## 5. Conclusion

In this paper, we studied surface damage of diamond exposed to low-energy high ion-flux hydrogen plasmas relevant for fusion applications. The chemical sputtering yield of diamond was measured while exposing high quality PECVD single crystals and polycrystalline coatings in Pilot-PSI to the ion flux of $(4-6) \cdot 10^{24}$ $D^+/m^2 s$ with impact energies of 7–9 eV per deuteron for a wide range of surface temperatures: $T_{surf} = 400-1200°C$. The spectroscopic relative measurements were put on an absolute scale by using a platinum mask deposited on one of the diamond samples. A bell-shaped dependence of the carbon sputtering yield vs. $T_{surf}$ was demonstrated, in agreement with a carbon erosion model of J. Roth [9]. The maximum carbon sputtering yield was observed at the surface temperature of about 900 °C. In contrast to experiments with thermal H atoms [7], chemical sputtering yields obtained in the present work are relatively high: $(0.5-1.4) \cdot 10^{-2}$ at/ion, close to those expected for graphite. The measured gross erosion yields were in good agreement with the carbon sputtering yields obtained by MD simulations under similar conditions. Our measurements also suggested that at relatively low surface temperature (400-550°C), the initial crystal orientation ([100] and [111]) of the sample does not demonstrate much difference in carbon erosion yield.

The XPS analysis suggested that the near surface region was modified during plasma exposure under our experimental conditions, which resulted in a ~1 nm thick hydrogenated amorphous carbon film. At surface temperatures less than 900 °C the $sp^2/sp^3$ ratio of the near surface region is increasing with increased surface temperature. MD simulation shows similar trend on the $sp^2/sp^3$ ratio, suggesting graphitization of the sub-surface region at 827 °C. At surface temperatures higher than 900 °C the polycrystalline sample was found to have similar $sp^2/sp^3$ ratio with that of unexposed single crystal sample. This could be the result of surface reconstruction caused by the dominant process of hydrogen desorption at such high temperature. The diamond surface state at high temperature could be interesting for application, although it is hard to achieve experimentally. At lower temperature it would be necessary to further reduce ion energy to limit defect creation. MD simulation was used to predict energy thresholds of carbon erosion with different surface temperatures. The near-surface sputtering of diamond shows threshold behaviour as a function of the $D^+$ impact energy; the chemical sputtering yield given by the model is below $3.3 \cdot 10^{-3}$ at/ion, provided that the $D^+$ impact energy is below 5 eV and $T_{surf} \leq 400°C$. With surface temperature ranging between 100°C and 400°C, diamond film can be a promising candidate for the production of negative-ion sources in NBI system where both ion flux and ion energy are lower ($10^{21}$–$10^{22}$ $m^{-2}s^{-1}$ and < 5 eV respectively) than in the present experimental conditions.

# Acknowledgements

This work was carried out within the framework of the French Research Federation for Fusion Studies (FR-FCM) and the EUROfusion Consortium and received funding from the Euratom research and training programme 2014-2018 under grant agreement No 633053. The views and opinions expressed herein do not necessarily reflect those of the European Commission. Financial support was also received from the French Research Agency (ANR) under grant 13-BS09-0017 H INDEX TRIPLED.

MD simulations provided by N. Ning and S. Khrapak were supported by the A*MIDEX project (Nr. ANR-11-IDEX-0001-02) funded by the French Government "Investissements d'Avenir" program managed by the French National Research Agency (ANR).